# Doubly Optimal Secure Multicasting: Hierarchical Hybrid Communication Network : Disaster Relief

G. Rama Murthy, Samdarshi Abhijeet , Deepti Singhal
International Institute of Information Technology
Hyderabad, India
E-mail: rammurthy@iiit.ac.in

*Abstract—* **Recently, the world has witnessed the increasing occurrence of disasters, some of natural origin and others caused by man. The intensity of the phenomenon that cause such disasters, the frequency in which they occur, the number of people affected and the material damage caused by them have been growing substantially. Disasters are defined as natural, technological, and human-initiated events that disrupt the normal functioning of the economy and society on a large scale. Areas where disasters have occurred bring many dangers to rescue teams and the communication network infrastructure is usually destroyed. To manage these hazards, different wireless technologies can be launched in the area of disaster. This paper discusses the innovative wireless technologies for Disaster Management. Specifically, issues related to the design of Hierarchical Hybrid Communication Network (arising in the communication network for disaster relief) are discussed.**

*Keywords-Disaster Management, Wireless Technology, Cognitive Radio, Wireless Sensor Network.*

## I. INTRODUCTION

Ever since the dawn of civilization, Homo sapiens civilization has been subjected to the wrath of disasters that are natural and man-made. These disasters involved loss of human life, property etc. Thus efforts were made to minimize loss to human life, property etc through the utilization of scientific, technological tools. The solutions involved preparedness for disaster, prediction or early warning of disaster, mitigation of loss after the disaster occurs. Disaster management is defined as encompassing mitigation, preparedness, response, and recovery efforts undertaken to reduce the impact of disasters, [1].

One important aspect of handling disaster is the ability to communicate the information over long distances and shorter time periods. The advent of electrical communication provided one major technological solution to handle disasters. The traditional solution to handle disasters was the wired communication network deployed over long distances.

Deployment of Internet provided one important technological solution to transmit, receive and share information. Thus, ideas to capitalize this communication infrastructure for disaster relief purposes are well underway. Also in recent times, "Cellular Wireless Network" pervaded the human life for wireless communication. Also, several wireless technologies based on innovative ideas are moving from laboratory stage to actual deployment. In this paper, we propose utilization of these wireless technologies to predict (early warning) and mitigate the effect of disasters.

*Cognitive Networking*

With the proliferation of wireless technologies, demand for electro-magnetic spectrum is increasing. In licensed band as well as unlicensed band, wireless services are contending for the spectrum. But it has been observed that over temporal and spatial dimension, the utilization of spectrum is not uniform i.e. even when the spectrum is allocated to Primary users, it is underutilized (over temporal and spatial dimensions). Hence researchers proposed the idea of cognitive networking whereby secondary users are allowed to use the spectrum when the primary user is not using it (or more broadly when the quality of service to the primary user is good enough). Thus, Dynamic Spectrum Access involves the utilization of cognitive radio parameters (such as centre frequency) can be dynamically tunable. In summary, cognitive networking involves:

- *Spectrum sensing:* Detecting unused spectrum and sharing the spectrum without causing hindrance to the primary users
- *Spectrum management:* Capturing the best available spectrum to meet user communication requirements.
- *Spectrum mobility:* Maintaining seamless communication requirements during the transition to better spectrum.
- *Spectrum sharing:* Providing the fair spectrum scheduling method among coexisting xG users.

This research paper is an effort to propose the design and performance analysis of wireless networks for disaster relief. The rest of the paper is organized as follows:

In Section 2, we summarize the wireless technologies that can be useful in disaster management/relief. In Section 3, some features / properties of such networks are summarized. Also a Hierarchical Hybrid Communication Network (HHCN) that can be used in disaster relief is discussed. Utilizing the results from information theory, secure communication among leader nodes (global and local) is summarized. An algorithm based on Huffman coding is proposed for secure message exchange in HHCN prior to actual communication (related to disaster relief). In Section4, energy efficient protocols for wireless networks (routing, fusion, localizations) are discussed. The paper concludes in Section 5.

## II. DISASTER RELIEF: COMMUNICATION NETWORKS

Across the world, wireless technology is fundamentally transforming the scope of possibility for disaster preparedness and response. In most disaster scenarios, different organizations have not been able to communicate with each other. This is because either the network becomes unavailable at some point in time, or different devices are not able to cooperate. Thus the design of communication architecture is very important for disaster management. This section discusses the common wireless communication architecture, available wireless technologies and the basic performance requirements of wireless network for disaster management.

Some of the wireless communication architectures that can be used for disaster management are as follows:
- Infrastructure based networks e.g cellular Nets, WLAN
- Pure Mobile Ad hoc Networks (MANETs)
- Hybrid Networks (Infrastructure based + MANETs)
- Hierarchical Hybrid Networks
- Mesh Networks

In Infrastructure based networks, there is one controlling node, called Base Station for some technologies and for some technologies access point, present for controlling the complete functioning of the network. The examples of such networks are cellular networks or wireless local area networks (WLANs). These networks are not completely suitable for disaster management because the complete functioning of network depends upon the controlling node, and if that node is not working complete communication will break.

*MANETS*

Mobile Ad hoc Network (MANET) is a collection of independent mobile nodes that can communicate to each other via wireless links. The union of these wireless links forms an arbitrary graph. The mobile nodes that are in radio range of each other can directly communicate, whereas others need the aid of intermediate nodes to route their packets. The nodes are free to move randomly and organize themselves arbitrarily; thus, the network's wireless topology may change rapidly and unpredictably. These networks are fully distributed, and can work at any place without the help of any infrastructure. This property makes these networks highly robust. Thus MANETs can be used for disaster management.

Hybrid of infrastructure based and mobile ad hoc network can also be used for disaster management based on the availability of network.

In hierarchical hybrid communication networks, nodes with different capabilities are used for different roles of network. Hierarchical hybrid communication network can be represented by tree or forest of trees data structure. **It should be noted that most of the organizations are structured based on the tree data structure, and several organizations constitute of forest of such trees**. Fault tolerant nature, low propagation delay, and diameter configurability of such structures make this architecture useful for communication scenarios. One example of hierarchical hybrid communication networks is shown in Fig. 1. Figure shows a graph representing three level hierarchical hybrid communication networks. At the maximum depth from the root node, the nodes are not connected with the nodes in same level, but for higher levels nodes are connected in same level also. **The hierarchical hybrid communication network architecture is suitable for disaster management.**

A mesh network is the network whose nodes are all connected to each other in a fully connected way. Mesh networks can be seen as one type of ad hoc network. The self-healing capability enables a fault tolerant network to operate when one node breaks down or a connection goes bad. As a result, the network is typically quite reliable, as there is often more than one path between a source and a destination in the network. Thus, this kind of network can also be used for disaster management.

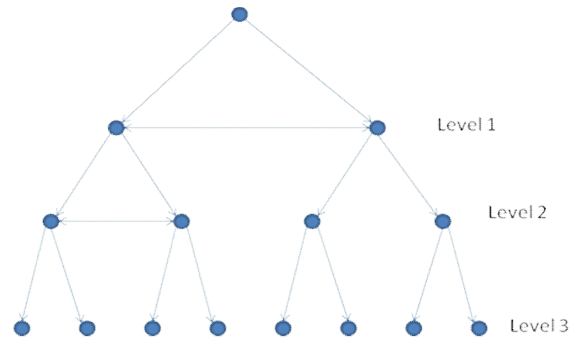

Figure 1. Example of Hierarchical Hybrid Communication Network

We now provide some light on the wireless technologies that can be used. The wireless technologies can be classified in two types,
- One inside building wireless networks. Wireless technologies such as Wireless Local Area Network (WLAN), Bluetooth, Wi-Fi, Wireless Sensor Networks (WSN) etc. can be used for inside building communication and
- Other as outside building wireless network. Wireless Wide Area Network (WWAN), Cellular Networks, and WSNs can be used outside building communication.

The criterion for selecting the wireless technology for disaster management depends on the requirements of such kind of networks. The disaster management network should provide real time communication, i.e. the delay of such networks should be very low. Disaster management network should be fault tolerant. It should be configurable on the fly and should adhere to other Quality of Service (QoS) requirements of communication networks.

## III. OPTIMAL SECURE MULTICASTING

As discussed in Section II, Wireless Sensor Networks can be used for disasters management. This paper considers a Cognitive Wireless Sensor Network (CWSN) and specifically hierarchical hybrid architecture of CWSN for disaster management. In most organizations that arise in physical reality, there is a hierarchical structure captured by a tree data structure. It can also be a forest formed by such trees. **One of the simplest possible such data structure is the balanced and unbalanced binary tree. Thus, it is clear that such fundamental data structure should be subjected to detailed study with respect to quantities of interest in applications such as disaster relief.**

In case of any communication network meant for handling disaster there are at least two possibilities, Network deployment prior to disaster, and Network deployment after disaster. Again these networks could be wired or wireless. These networks should be designed to provide

a) Desired quality of service
b) Fault tolerance
c) Ease of deployment
d) Re-configurability.

We now focus our attention on a common feature that arises in Hierarchical Hybrid Communication Networks (HHCN) applications. For concreteness, consider a balanced binary tree. Note that the feature arises in other trees modelling HHCN. Nodes at the highest depth are not at all connected to one another. They are only connected to their immediate predecessor. This predecessor is one level above the leaf nodes. Further theses immediate predecessor are relatively better connected to one another. Most of the time, these nodes are not fully connected. As we traverse the tree from the leaf nodes to the root node, the connectivity structure becomes denser and number of nodes becomes lesser. The example of this type of data structure is illustrated in Figure 1.

Let us assume that network architecture is represented by a D-ary tree of maximum depth $n_{max}$. Total number of nodes at depth $n_j$ is given by $D^{n_j}$. Let's assume that $s_j$ nodes out of total $D^{n_j}$ nodes are elected as local leaders, then the probability that a single node is chosen as local leader is given by

$$P\{a\ node\ is\ chosen\ as\ local\ leader\} = \frac{s_j}{D^{n_j}} = t_j$$

Thus from all possible nodes in HHCN, some are chosen as the leader nodes. We are interested in the following quantities in the design of HHCN:

- Probability that a randomly chosen node is a leader at level $j$

$$= \left(\frac{D^{n_j}}{D^{n_{max}+1} - 1}\right)\left(\frac{s_j}{D^{n_j}}\right) = \left(\frac{s_j}{D^{n_{max}+1} - 1}\right)$$

- Probability that a randomly chosen node is a local leader

$$= \frac{\sum_{i=1}^{M} s_j}{(D^{n_{max}+1} - 1)}$$

Where $n_{max}$ is total number of levels in D-ary tree.

*Hierarchial Hybrid Communication Network (HHCN): Information Theory*

Now let us consider the design of HHCN arising in disaster communication network (as well as networks arising in the hierarchies of various organisations). We have the following design issues:-

- The HHCN is represented as a D-ary tree with the root node as global leader.
- Some local leaders are chosen as leaders at various depths. Path from global leader to local leaders are "prefix" free i.e. the path to a leader at larger depth is not a "prefix" ( descendent path ) of a path to a local leader at lesser depth.
- Local leaders are assigned hierarchical importance, captured through probability Pj at depth "j". Thus } $p_1, p_2, \ldots, p_{n\ max}$} is the probability mass function representing importance of local leaders at various depths. Also let

  $n_i$ : Depth of communication path to local leaders from root node (i.e. global leader )

Thus in summary, X is the random variable assuming various depths from root node.

Intuition : If the hierarchical importance of a local leader is low, assign him path of large length from root node in the D-ary tree conversely, if a leader is highly important, assign him a path of small length from root node in the D-ary tree.

Problem statement : The paths to "Local Leaders" from the "Global Leader" (root node) should be "Prefix free " and at the same time the average length of path to local leaders should be as small as possible.

Goal : Transfer results from "Source Coding" to OPTIMALLY choose leaders in the hierarchy represented by the D-ary tree corresponding to HHCN.

**Now we know the probability that a randomly chosen node is a local leader**. Thus these local leaders can be seen as a codeword like in case of source coding of information theory in a D-ary local leader. These elected local leaders should be at Prefix free path in a D-ary tree for secured communication property. If these local leaders are not at prefix free path, then the message forwarded from global leaders (BS) will propagate to the nodes at depth "j" and also at depths strictly larger than "j". From security considerations, the initial messages from a global leader to a local leader at depth "j"can be heard by local leaders at depth lesser than "j", but should not be heard by local leaders at depth greater than "j". And same as the information theory case where codeword should be uniquely decodable, the path to the local leaders should be Prefix free in this case. As $n_1, n_2, \ldots, n_M$ are also the lengths from global leader to the local leaders, in terms of the hop count, all the paths are Prefix free in a D-ary tree if

$$\sum_{i=1}^{M} D^{-n_i} \leq 1$$

This result is known as Kraft's Inequality in information theory. Thus if the lengths of the local leaders are chosen such that Kraft's Inequality holds true, then there exists a Prefix free path from the global leader to all local leaders.

**Lemma:** If $n_i$'s are integer values with the following relation, then kraft's inequality is true.
$$n_2 = n_1 + 1, n_3 = n_2 + 1 = n_1 + 2, \cdots$$
$$\cdots, n_M = n_{M-1} + 1 = n_1 + M - 1$$

**Proof:** We can write the left hand side of the inequality as
$$\sum_{i=1}^{M} D^{-n_i} = D^{-n_1} + D^{-(n_1+1)} + D^{-(n_1+2)} + \cdots + D^{-(n_1+M-1)}$$
$$= D^{-n_1}[1 + D^{-1} + D^{-2} + \cdots + D^{-(M-1)}]$$

Let $S = 1 + D^{-1} + D^{-2} + \cdots + D^{-(M-1)}$, then
$$D^{-1}S = D^{-1} + D^{-2} + \cdots + D^{-(M-1)} + D^{-M}$$
$$(D^{-1} - 1)S = D^{-M} - 1$$
$$S = \frac{D^{-M} - 1}{D^{-1} - 1}$$

Thus,
$$\sum_{i=1}^{M} D^{-n_i} = D^{-n_1}\left(\frac{D^{-M} - 1}{D^{-1} - 1}\right)$$

Now let us consider the very special case of $D = 2$.
$$\sum_{i=1}^{M} 2^{-n_i} = 2^{-n_1}\left(\frac{2^{-M} - 1}{2^{-1} - 1}\right)$$
$$= 2^{-n_1}\left(\frac{1 - 2^{-M}}{\frac{1}{2}}\right)$$
$$= 2^{-n_1+1}(1 - 2^{-M})$$

For $n_1 = 1$,
$$\sum_{i=1}^{M} 2^{-n_i} = \left(1 - \frac{1}{2^M}\right)$$
$$\sum_{i=1}^{M} 2^{-n_i} \leq 1, \forall M$$

Similarly Kraft's inequality can also be proved for other values of $D$ and $n_i$. □.

As in the case of source coding, Huffman coding type algorithm can be easily designed to arrive at Prefix free paths to local leaders. Let the nodes at depth "j" be designated as level 'j' i.e. Hop count of the node from root node.

Suppose each link fails with probability $q$ and the link failure on different levels are independent. Then the following calculations for the reliability of communication can be made:
- Probability that a communication path exists from root node to local leader at level $n_j$
$$= (1-q)^{n_j}, \quad 1 \leq n_j \leq M$$
- Probability that $([n]_j - 1)$ links are reliable and last link to the leader fails
$$= (1-q)^{n_j-1}q, \quad 1 \leq n_j \leq M$$

In a general case there could be more than one local leader at one level, same as source coding. A relatively better model is superposition or merging of D-ary trees with different value of D. For example binary tree merged with trinary tree as shown in Fig. 2. Note: The above results are being generalized to arbitrary graphs

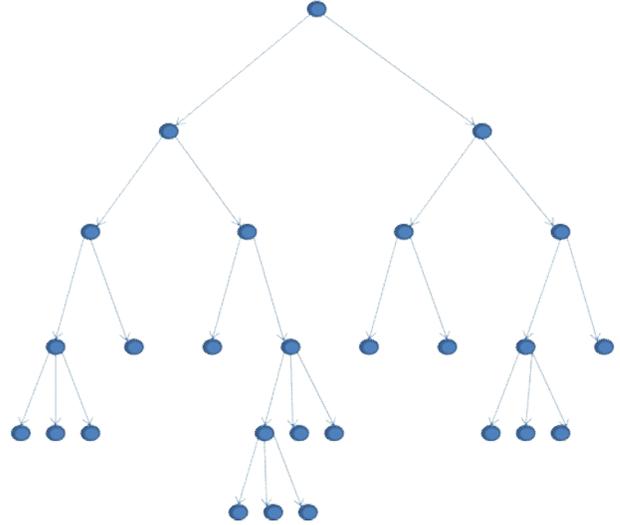

Figure 2. BinaryTree merged with Trinary Tree

Summary ( So Far ):

In the above discussion, we have discussed how results from source coding ( in Information Theory ) can be utilized to find optimal ( in the sense of minimizing the mean depth from root node ) "prefix free paths" from global leader to the local leaders ( of varying importance captured through probabilities ). The discussion assumed that the hierarchical hybrid communication network is abstracted through a D-ary tree. Specifically prefix free paths are derived using huffman coding type algorithm.

Doubly Optimal Secure Multicasting:

Now we interpret the above results from the point of view of multicasting from a chosen root node ( global leader ) to selected nodes ( local leaders ) of varying importance captured through probabilities.

It should be noted that the constraint of finding prefix free paths ensures security of communication ( for sharing, say, cryptographic keys ). Thus, we solved the problem of OPTIMAL ( in a well defined/useful sense ) SECURE MULTICASTING in a D-ary tree (e.g. bibnary tree ).

Now we generalize the above result to arbitrary connected graph representing a communication network

- Suppose the edges of the connected graph are associated with some weights ( representing say "delay" ). From the chosen root node, one can find spanning tree of the graph and more importantly the MINIMUM SPANNING TREE ( optimal in the well defined sense ) of the graph

- Suppose the degree of all the vertices ( in the minimum spanning tree ) is atleast TWO ( may be more ) except for the leaf nodes of the MINIMUM SPANNING TREE. Thus the spanning tree has atleast a binary tree imbedded in it.

- Now as in the above case, solve the problem of finding optimal prefix free paths ( using Huffman Coding type algorithm ) in the imbedded binary tree. Thus, we solved the problem of DOUBLY OPTIMAL SECURE MULTICASTING in a structured graph.

- Currently, we are generalizing the above results to arbitrary graphs.

## IV. PROTOCOLS IN WIRELESS NETWORKS

This section discusses some of the protocol details about the architecture discussed in Section III.

### A. Routing

In the hierarchical hybrid communication network architecture, the local leaders are elected based on the Kraft's inequality condition. The other nodes form the cluster like in LEACH [2] or HEAD [3] assuming those local leaders as cluster heads. Routing between clusters is done by Level Controlled Gossiping Scheme [4]. In Level controlled gossip, the probabilities associated with each level can be set during leveling phase. The probabilities decrease as we move from inner levels to outer levels as shown by the relation.

$$P_1 > P_2 > P_3 > ... > P_{n-1} > P_n.$$

$P_j$ represents the transmission probability from node at level 'j'. Here $P_1, P_2, P_3 ... P_n$ denote the transmission probabilities with which a node should forward a received message to the nodes in other levels. These probabilities denote these probabilities can be varied any time by the base station to suite the monitoring requirements. When an event is detected the message is broadcast with the probability of that level. Once a message is received by a node, it checks to see if it is from a higher level. If it is from a lower or same level the message is discarded. When a node in the lower level receives this message from higher level, it transmits the message with the probability of its corresponding level. So, the same event is being transmitted at lower probabilities in outer layers and higher probabilities in inner layers. The advantage of level controlled gossip is, it balances the gossip (probabilistic flooding) happening in the levels according to the proximity of the level to the base station. This approach balances the network life time and monitoring reliability.

Routing can also be done using Leveling Sectoring approach discussed in [5], if sectoring is possible in the area.

### B. Localization

Since most of the disaster management applications depend on a successful localization, i.e. to compute their positions in some coordinate system, it is of great importance to design efficient localization algorithms. Unfortunately, for a large number of sensor nodes, straightforward solution of adding GPS to all nodes in the network is not feasible because:

- In the presence of dense forests, mountains or other obstacles that block the line-of-sight from GPS satellites, GPS cannot be implemented.
- The power consumption of GPS will reduce the battery life of the sensor nodes and also reduce the effective lifetime of the entire network.
- In a network with large number of nodes, the production cost factor of GPS is an important issue.
- Sensor nodes are required to be small. But the size of GPS and its antenna increases the sensor node form factor.

For these reasons an alternate solution of GPS is required which is cost effective, rapidly deployable and can operate in diverse environments. Localization can also be done using uses the leveling - sectoring based localization [6] where each Cluster head is identified by two coordinates, i.e. level id and sector id. First the entire cognitive network area is divided into various levels as discussed. To differentiate between the nodes within the same level, network field is divided into equiangular regions called sectors as [5]. Each sector is uniquely identified with the help of sector ID. In order to inform their sector IDs to nodes, the Base station sends Sector Broadcast Packets (SBP) with directional antennas such that only nodes within a sector receive this broadcast. Thus nodes know their location in terms of $(L_i, \theta_i)$, where $L_i$ is the level id of node and $\theta_i$ is the sector id of the node.

### C. Fusion

Information fusion deals with the combination of information from same source or different sources to obtain improved fused estimate with greater quality or greater relevance. As larger amount of sensors are deployed in harsher environment, it is important that sensor fusion techniques are robust and fault-tolerant, so that they can handle uncertainty and faulty sensor readouts.

Let $I_1, I_2, ... , I_n$ be the interval estimates from n abstract sensors, and maximum f of them could be faulty. Four functions were developed representing four milestones in this area discussed in [7], these four functions are shown in Fig.3.

M function [8] is defined as the smallest interval that contains all the intersections of $(n - f)$ intervals. It is guaranteed to contain the true value provided the number of faulty sensors is at most f, i.e. $f_{max} = f$. However, M function exhibits an unstable behavior in the sense that a slight difference in the input may produce a quite different output. This behavior was formalized as violating Lipschitz condition [9].

The $\Omega$ function [10] is also called the overlap function. $\Omega(x)$ gives the number of intervals overlapping at x. $\Omega$ function results in an integration interval with the highest peak and the widest spread at a certain resolution. The $\Omega$ function is also robust, satisfying Lipschitz condition, which ensures that minor changes in the input intervals cause only minor changes in the integrated result.

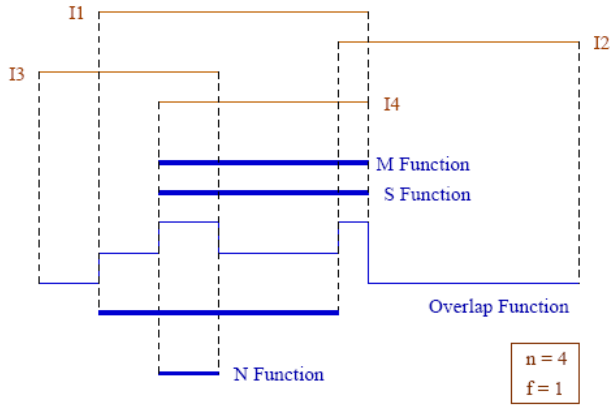

Figure 3. Fusion Functions

The N function [11] improves the $\Omega$ function to only generate the interval with the overlap function ranges *[n − f, n]*. It also satisfies Lipschitz condition.

S function of Schmid and Schossmaier [12] returns a closed interval *[a, b]* where *a* is the *(f + 1)$^{th}$* maximum left end point and *b* is the *(f + 1)$^{th}$* minimum right end point of the intervals i.e. there are exactly *f* left end points to the right of *a* when the left end points are sorted in increasing order and similarly there are *f* right end points to the left of *b* when right end points are sorted in increasing order. This function also satisfies the lipschitz condition [12]. Schimd et al also presented that S function is an optimal function from the listed functions.

## V. CONCLUSIONS

In this research paper, utilization of currently popular wireless networks for disaster relief is summarized. A Hierarchical Hybrid Communication Network (HHCN) is proposed to model hierarchical organization of the wireless communication network for disaster relief. Secure message exchange algorithm in HHCN is proposed. Energy efficient protocols in wireless networks are proposed.